\documentclass[%
 reprint,
superscriptaddress,
 amsmath,amssymb,
 aps,
floatfix,
]{revtex4-1}

\usepackage{graphicx}
\usepackage{dcolumn}
\usepackage{bm}

\usepackage{gensymb}
\usepackage{float}

\usepackage{graphicx}
\usepackage[english]{babel}
\usepackage[utf8]{inputenc}

\usepackage[version=4]{mhchem}

\usepackage{xcolor}

\begin{document}
\title{Circular photogalvanic effect in Cu/Bi bilayers}

\author{Hana Hirose}
\affiliation{Department of Physics, The University of Tokyo, Tokyo 113-0033, Japan}

\author{Naoto Ito}
\affiliation{Department of Physics, The University of Tokyo, Tokyo 113-0033, Japan}

\author{Masashi Kawaguchi}
\affiliation{Department of Physics, The University of Tokyo, Tokyo 113-0033, Japan}

\author{Yong-Chang Lau}
\affiliation{Department of Physics, The University of Tokyo, Tokyo 113-0033, Japan}
\affiliation{National Institute for Materials Science, Tsukuba 305-0047, Japan}

\author{Masamitsu Hayashi}
\affiliation{Department of Physics, The University of Tokyo, Tokyo 113-0033, Japan}
\affiliation{National Institute for Materials Science, Tsukuba 305-0047, Japan}



\date{\today}

\begin{abstract}
We have studied the circular photogalvanic effect (CPGE) in Cu/Bi bilayers. When a circularly polarized light in the visible range is irradiated to the bilayer from an oblique incidence, we find a photocurrent that depends on the helicity of light. Such photocurrent appears in a direction perpendicular to the light plane of incidence but is absent in the parallel configuration. 
The helicity dependent photocurrent is significantly reduced for a Bi single layer film and the effect is nearly absent for a Cu single layer film.
Conventional interpretation of the CPGE suggests the existence of spin--momentum locked band(s) of a Rashba type in the Cu/Bi bilayer. 
In contrast to previous reports on the CPGE studied in other systems, however, the light energy used here to excite the carriers is much larger than the band gap of Bi. Moreover, the CPGE of the Cu/Bi bilayer is larger when the energy of the light is larger: the helicity dependent photocurrent excited with a blue light is nearly two times larger than that of a red light.
We therefore consider the CPGE of the Cu/Bi bilayer may have a different origin compared to conventional systems.
\end{abstract}

\maketitle
Spin--momentum locked bands are one of the key signatures of the emergence of topologically protected states in topological insulators and Weyl semimetals\cite{Hasan_2010_RMP}. Such bands also appear in heterostructures with broken structure inversion symmetry and/or large spin orbit coupling (SOC)\cite{RojasSanchez_2014_NatComm,Lesne_2016_NatMater,Song_2017_SciAdv}. The spin texture of the spin--momentum locked bands in the reciprocal space depends on the symmetry of the system. For example, the electron's spin and momentum directions are orthogonal to each other for systems that can be described by a Rashba Hamiltonian\cite{Bychkov_1984_JETP,Manchon_2015_NatMater}. 

The presence of spin--momentum locked bands within the bulk or at surfaces/interfaces allows generation of non-equilibrium spin accumulation when current is passed to the system\cite{Edelstein_1990_SSC,Kato_2004_PRL}. Significant effort has been placed to generate spin accumulation in semiconductor heterostructures\cite{Nitta_1997_PRL,Zutic_2004_RMP}. Recent studies have extended such effort into metallic heterostructures\cite{RojasSanchez_2014_NatComm,Manchon_2015_NatMater,Isasa_2016_PRB,Matsushima_2017_APL}, where the SOC can be larger than that of typical semiconductor heterostructure constituents. Current-induced spin accumulation at interfaces has been reported in metallic heterostructures which manifests itself in magnetization switching and domain wall motion\cite{Miron_2011_NatMater,Miron_2011_Nature}. It is thus of high importance to identify the presence of spin--momentum locked bands in thin film heterostructures. 

Angle resolved photoemission spectroscopy (ARPES) is a powerful tool to study band structures and has been used to reveal the surface electronic states of, for example, topological insulators and Weyl semimetals\cite{Hasan_2010_RMP}. However, its use is typically limited to clean surfaces and involves difficulty in studying interface states of films which are not particularly clean (e.g. films deposited by sputtering). To study spin--momentum locking of such interface states, it has been shown recently that combination of spin pumping and the inverse Rashba-Edelstein effect (IREE) allows its direct probing\cite{RojasSanchez_2014_NatComm,Lesne_2016_NatMater,Song_2017_SciAdv}. Spin current generated from a ferromagnetic layer via spin pumping diffuses into an interface with strong spin orbit coupling where IREE converts spin current into electric current. The Rashba parameter of interface states has been estimated using this approach and is expanding its use to different systems.   

Another approach to studying the electronic structure is the circular photogalvanic effect (CPGE)\cite{Ganichev_2002_Nature,Ivchenko_2017_condmat}.
When a circularly polarized light is irradiated to a sample from an oblique incidence, carriers with fixed spin orientation are excited due to the selection rule (under the presence of spin orbit coupling) and conservation of spin angular momentum. The excited carriers diffuse along a direction perpendicular to the spin orientation when a spin--momentum locked band of Rashba type is involved in the excitation process. Under such circumstance, a spin polarized photocurrent flows orthogonal to the light plane of incidence. Studies of such helicity dependent anisotropic photocurrent allows direct probing of the spin--momentum locked bands within the sample. The CPGE has been observed in semiconductor heterostructures\cite{Ganichev_2004_PRL,Weber_2005_APL,Kohda_2015_APL} and more recently in topological insulators\cite{Hosur_2011_PRB,Kastl_2015_NatComm,Okada_2016_PRB,Pan_2017_NatComm}, Weyl semimetals \cite{Ma_2017_NatPhys} and the two dimensional calchogenides\cite{McIver_2012_NatNano,Yuan_2014_NatNano}. 
In contrast to the spin voltaic effect that has been observed in semiconductor heterostructures due to optically induced spin accumulation\cite{Zutic_2002_PRL,Zutic_2006_PRL,Kondo_2006_JJAP,Endres_2013_NatComm}, the CPGE does not require any magnetic field or ferromagnetic layer. In addition, the photovoltage generated from the spin voltaic effect appears along the film (interface) normal whereas the photocurrent associated with the CPGE is often observed along the film plane. 

\begin{figure}[t]
\centering
\includegraphics[scale=0.24]{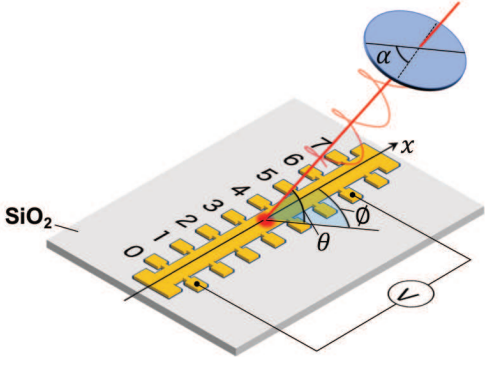}
\caption{Schematic illustration of the experimental setup. The yellow caterpillar-like structure illustrates the area where the film is deposited through a shadow mask: the surrounding grey area represents the substrate. Light is irradiated from an oblique angle $\theta$ with respect to the film plane. $\phi$ is defined as the angle between the light plane of incidence and the long axis of the wire. }
\label{fig:geometry}
\end{figure}

Here we use the CPGE to study the presence of spin--momentum locked bands in (semi-)metallic heterostructures. In Cu/Bi bilayers, we measure the photocurrent when a polarized light in the visible range (405 nm and 635 nm) is irradiated to the sample from an oblique angle. Light helicity dependent photocurrent is found along the direction orthogonal to the light plane of incidence but is negligibly small in the direction parallel to it. The helicity dependent anisotropic photocurrent is significantly reduced in a Bi single layer film and is nearly zero for a Cu single layer film. 
Although the CPGE may originate from spin--momentum locked bands formed at local Cu/Bi interfaces, we infer it may include contributions from the bulk of the bilayer as the light energy is much larger than the band gap of Bi.

Films are deposited via RF magnetron sputtering on Si substrates coated with 600 nm thick thermal oxide (SiO$_2$). We show representative results from three film structures, A: sub./0.5 Ta/2 Cu/10 Bi/2 MgO/1 Ta, B: sub./0.5 Ta/2 Cu/2 MgO/1 Ta and C: sub./10 Bi/2 MgO/1 Ta (thickness in nm). The 2 MgO/1 Ta is used as a capping layer to prevent oxidation and possible subsequent degradation of the top layer. The half a nanometer thick Ta layer underneath the 2 nm Cu layer (samples A and B) is used as a seed layer to promote uniform growth of the Cu layer. The resistivities of the Bi and Cu layers are measured using four-point probe technique and are found to be $\sim 820 \ \mu \Omega \cdot$cm and $\sim 110 \ \mu \Omega \cdot$cm, respectively. 
Films are deposited through a metallic shadow mask to create patterned structures. A schematic illustration of the patterned structure is shown in Fig.~\ref{fig:geometry}. An elongated wire with contact pads attached to the sides and ends is used to study the spatial profile of the photocurrent/photovoltage. The width of the wire is $\sim$ 0.4 mm. The side contact pads are located $\sim$ 1 mm apart and are labeled with numbers (0 to 7) as depicted in Fig.~\ref{fig:geometry}. 

A continuous wave (CW) semiconductor laser (wavelength: 405 nm (635 nm), power: 4.0 mW (4.5 mW)) is used to generate light. The laser light is chopped at a frequency $f \sim$ 311 Hz: the frequency is chosen to avoid noise from the power line. The laser spot at the sample is a circle with a diameter of $\sim$ 0.5 mm. The light is irradiated at the center of the sample, between side contacts 3 and 4, from an oblique angle $\theta$ (see Fig.~\ref{fig:geometry} for the definition of angles $\theta$ and $\phi$). We use $\theta \sim 45^{\circ}$ which gives a relatively large CPGE\cite{Weber_2005_APL,Okada_2016_PRB,Yuan_2014_NatNano}. The azimuthal angle $\phi$ is set to either 0$^{\circ}$ or 90$^{\circ}$ to study the relationship between the light plane of incidence and the direction of the photocurrent/photovoltage. The light plane of incidence is defined as the plane that contains both the film normal and the light propagation vector. 

\begin{figure}[t]
\centering
\includegraphics[scale=0.34]{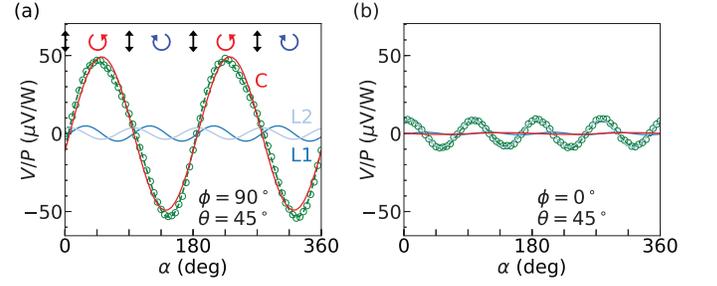}
\caption{Normalized photovoltage ($V/P$) of Cu/Bi bilayer (sample A). Voltage between side contacts 7 and 0 are measured. 
The light (wavelength: 405 nm) is incident from $\phi=90^{\circ}$ (a) and $\phi=0^{\circ}$ (b). The open circles represent experimental data, the green dashed line shows fit to the data with Eq.(\ref{eq:fitting}). The red, dark blue and light blue solid lines show contributions from the $C$, $L_1$, and $L_2$ terms, respectively. The constant $D$ term is subtracted from the results. 
}
\label{fig:CuBi_raw}
\end{figure}

We use an open circuit condition to measure the photovoltage $V$ and estimate the corresponding photocurrent by dividing $V$ with the resistance of wire between contacts 3 and 4. 
The resistance is determined using the four-point probe technique. 
The photovoltage $V$ between side contacts $i$ ($i=0...7$) and 0 (contact 0 is always set as the reference) is measured using a lock-in amplifier with its frequency locked to the laser light chopping frequency $f$. The reflected light from the sample is measured using a photodetector and its signal is measured together with the photovoltage to determine the phase of $V$ with respect to the light irradiation.

A quarter wave plate is used to control the light helicity. We rotate the quarter wave plate within the plane normal to the light propagation. The angle of the rotation is defined as $\alpha$ (see Fig.~\ref{fig:geometry}). Here the light is linearly (s-) polarized when $\alpha=0^{\circ}$, 90$^{\circ}$, $180^{\circ}$, $270^{\circ}$. The light is circularly polarized when $\alpha=45^{\circ}$, 225$^{\circ}$ (left handed) and 135$^{\circ}$, 315$^{\circ}$ (right handed). At a given $\theta$ and $\phi$ ($\theta$ is always fixed to 45$^{\circ}$), the quarter wave plate is rotated to measure the $\alpha$ dependence of $V$.  All measurements are performed at room temperature under ambient condition. 

The open circles in Fig.~\ref{fig:CuBi_raw} show the light helicity dependence of photovoltage $V$, normalized by the laser power $P$, between side contacts 7 and 0 of the Cu/Bi bilayer (sample A). The light plane of incidence is orthogonal ($\phi \sim 90^{\circ}$) and parallel ($\phi \sim 0^{\circ}$) to the wire's long axis in Figs.~\ref{fig:CuBi_raw}(a) and \ref{fig:CuBi_raw}(b), respectively. For $\phi \sim 90^{\circ}$, we find a large difference in $V/P$ when left and right handed circularly polarized light is irradiated. In contrast, $V/P$ for $\phi \sim 0^{\circ}$ shows little dependence on the light helicity and instead a difference in $V/P$ is observed when circularly polarized light and linearly polarized light are irradiated.

\begin{figure}[t]
\centering
\includegraphics[scale=0.35]{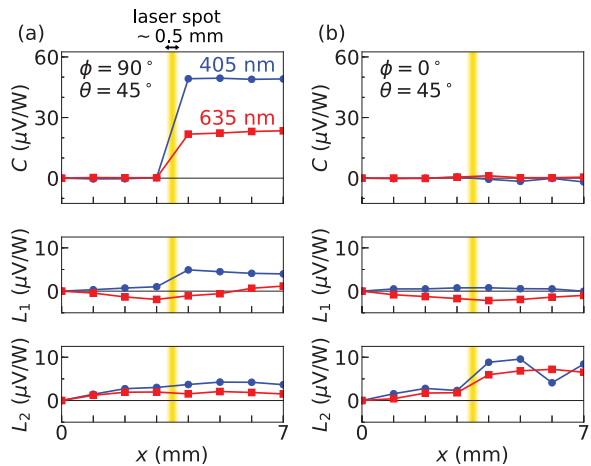}
\caption{Spatial profile of $C$ (top panels), $L_1$ (middle panels) and $L_2$ (bottom panels) for the Cu/Bi bilayer (sample A). The light is incident from $\phi=90^{\circ}$ (a) and $\phi=0^{\circ}$ (b).
The blue circles and the red squares show the corresponding parameters ($C$, $L_1$ and $L_2$) obtained from measurements of the photovoltage using laser light with wavelength of 405 nm and 635 nm, respectively. The yellow shaded area represents the location of the laser spot.}
\label{fig:location}
\end{figure}

To characterize the normalized photovoltage, we fit the experimental results with the following phenomenological form:
\begin{equation}
\begin{aligned}
    V/P=&C\sin{2(\alpha + \alpha_0)}\\
    &+L_1\sin{4(\alpha + \alpha_0)}+L_2\cos{4(\alpha + \alpha_0)}+D
\label{eq:fitting}
\end{aligned}
\end{equation}
where $C$ represents the difference in the photovoltage when left and right handed polarized light are irradiated, $L_1$ and $L_2$ are the change in the photovoltage between irradiation of circularly polarized light and linearly polarized light (with a phase difference of 22.5$^{\circ}$ between the two). In theory, $C$ represents the strength of the CPGE and $L_1$ and $L_2$ correspond to the linear photogalvanic and the photon-drag effects, respectively\cite{McIver_2012_NatNano,Okada_2016_PRB,Olbrich_2014_PRL}. $D$ is a constant independent of $\alpha$. $\alpha_0$ is an offset angle associated with the quarter wave plate and the experimental setup: here $\alpha_0 \sim -7^{\circ}$ and $\sim 3^{\circ}$ for the setup with laser wavelength of 405 nm and 635 nm, respectively.

The green dashed lines in Figs.~\ref{fig:CuBi_raw}(a) and \ref{fig:CuBi_raw}(b) show the results of fitting with Eq. (\ref{eq:fitting}): the solid lines show contributions from each component ($C$, $L_1$, $L_2$). As evident, the CPGE term is dominant (solid red line) for $\phi \sim 90^{\circ}$ (Fig.~\ref{fig:CuBi_raw}(a)): $C$ is $\sim 49$ $\mu$V/W which corresponds to a photocurrent of $\sim 80$ nA/W. This is in similar magnitude with what has been observed in other systems\cite{Yuan_2014_NatNano,Okada_2016_PRB}. In contrast, $C$ is nearly zero when $\phi \sim 0^{\circ}$ and we find the $L_2$ term dominates the signal with $L_2 \sim 9$ $\mu$V/W. 

The spatial profile of each component of the normalized photovoltage is shown in Fig.~\ref{fig:location}. The position of one of the voltage probes is varied from side contacts 0 to 7: the reference voltage probe is fixed to contact 0. Results from light wavelength of 405 nm (blue circles) and 635 nm (red squares) are shown together. We find nearly zero $C$ for $\phi \sim 0^{\circ}$, consistent with the results of Fig.~\ref{fig:CuBi_raw}(b) . 
For $\phi \sim 90^{\circ}$, $C$ abruptly changes across the laser light spot, i.e. between contacts 3 and 4. $C$ is constant on both sides of the light spot. 
These results suggest that the CPGE occurs within the area of side contacts 3 and 4 (it is likely that the effect takes place within the laser spot).
As the excited carriers due to the CPGE flow within this area, we divide the measured photovoltage $V$ by the resistance of the wire between contacts 3 and 4 to estimate the corresponding photocurrent $I$ from $V$ (note that $V$ is constant across the laser spot). 
$I$ will thus provide the lower limit of the actual photocurrent generated via the CPGE.

We find $C$ is larger for the blue light (wavelength: 405 nm) than that for the red light (wavelength: 635 nm). Note that energy of the blue and red light is much larger than the band gap of Bi\cite{Liu_1995_PRB,Lenoir_1996_JPCS}: it is thus not obvious whether the number of excitation that contributes to the generation of the CPGE scales with the light energy in the visible range.  
We find also a step like features across the laser spot in $L_1$ with $\phi \sim 90^{\circ}$ and $L_2$ with $\phi \sim 0^{\circ}$. In addition, $L_1$ and $L_2$ show a broad background signal with an extremum near the laser spot. The tail of the background signal extends to the edge of the wire, and the profile looks similar for both configurations $\phi \sim 90^{\circ}$ and $0^{\circ}$. Such large length scale (millimeter long diffusion length) is consistent with laser-induced heating and thermal diffusion. 

\begin{figure}[t]
\centering
\includegraphics[scale=0.23]{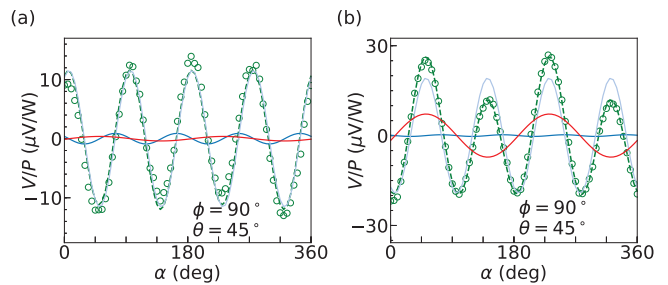}
\caption{(a,b) Normalized photovoltage ($V/P$) of the Cu single layer film, sample B (a) and the Bi single layer film, sample C (b). Voltage between side contacts 7 and 0 are measured. 
The light (wavelength: 405 nm) is incident from $\phi=90^{\circ}$. The open circles represent experimental data, the green dashed line shows fit to the data with Eq.(\ref{eq:fitting}). The red, dark blue and light blue solid lines show contributions from the $C$, $L_1$, and $L_2$ terms, respectively. The constant $D$ term is subtracted from the results. 
}
\label{fig:refsamples}
\end{figure}

To identify the origin of the CPGE in Cu/Bi bilayer, we have studied the photovoltage of the Cu (sample B) and Bi (sample C) single layers. The $\alpha$ dependence of $V/P$ for the two samples is shown in Fig.~\ref{fig:refsamples}. Side contacts 7 and 0 are used to probe the photovoltage. Blue light (wavelength: 405 nm) is irradiated and the angle $\phi$ is set to $90^{\circ}$. The experimental results in Fig.~\ref{fig:refsamples} are fitted using Eq. (\ref{eq:fitting}): the fitting results are displayed using the dashed and solid lines. A summary of the fitting parameters for all samples are shown in Table \ref{table:data}.

\renewcommand{\arraystretch}{1.5}
\begin{table}[t]
  \caption{Photogalvanic constants, $C$, $L_1$ and $L_2$, for samples A, B and C. The normalized photocurrent $I/P$ [nA/W] is obtained by dividing the normalized photovoltage $V/P$ [$\mu$V/W] with the measured resistance of the wire between side contacts 3 and 4. The photovoltages of side contacts 7 and 0 are taken: the light wavelength is 405 nm.\\}
  \label{table:data}
  \centering
  \begin{tabular}{lr|@{\hspace{0.5cm}}rr@{\hspace{0.5cm}}rr@{\hspace{0.5cm}}rr}
    \hline
    Sample  & $\phi$  & \multicolumn{2}{c}{$C$} &\multicolumn{2}{c}{$L_1$} & \multicolumn{2}{c}{$L_2$}  \\
       &   & $V/P$ &$I/P$ &$V/P$ &$I/P$ & $V/P$ &$I/P$  \rule[-1.2mm]{0mm}{5mm}\\
    \hline \hline
    A: Cu/Bi  & $90^{\circ}$  & 49.2  &80.0 & 4.9 &8.0  & 3.7 &6.0 \\
      & $0^{\circ}$   & $-0.5$ & $-0.8$  & 0.8 & 1.3& 8.8 &14.3\\
    \hline 
    B: Cu  & $90^{\circ}$  & 0.4   & 0.3 & $-0.9$ &$-0.6$ & 11.6 &8.2 \\
    \hline 
    C: Bi  &$90^{\circ}$  &  7.2  & 5.0 & $-0.3$ & $-0.2$& $-19.2$ &$-13.5$ \\
    \hline\hline 
  \end{tabular}
\end{table}

For the Cu single layer (Fig.~\ref{fig:refsamples}(a)), we find $V/P$ that is dominated by the $L_2$ term: $L_1$ and $C$ are nearly zero. Table \ref{table:data} shows that the $L_2$ term is similar in magnitude for all samples. This is consistent with the observation that the $L_2$ term originates from an effect that occurs within the bulk of the film\cite{McIver_2012_NatNano}.
(Note that the degree of absorption is different when the irradiated light is linearly polarized and circularly polarized. Such difference likely contributes to the generation of $L_2$.) 
The CPGE of the Bi single layer film is non-negligible: we find $C \sim 7 \mu$V/W (the corresponding photocurrent is $\sim$ 5 nA/W).  Although Bi itself is known to possess spin-momentum locked bands in its surface state\cite{Koroteev_2004_PRL,Marcano_2010_PRB}, the observed $C$ is nearly an order of magnitude smaller than that found in the Cu/Bi bilayer. 
These results support the conventional interpretation that the CPGE of the Cu/Bi bilayer (sample A) originates, if not entirely, from spin--momentum locked bands of Rashba type at the Cu/Bi interface.

Atomic force microscopy (AFM) imaging and X-ray diffraction (XRD) are used to characterize the film structure (see supplementary material). The AFM images show the Cu/Bi bilayer (sample A) and the Bi single layer film (sample C) grow in three dimensional form\cite{Missana_1996_APA} with typical grain size of $\sim100$ nm and $\sim$50 nm, respectively. From the XRD spectrum, we find the grains of the bilayer are textured but oriented in random directions. As we find diffraction peaks from the plane perpendicular to the c-axis of the hexagonal phase of Bi, which is known to give relatively large Rashba coupling at surfaces or interfaces between Bi(001) and various materials\cite{RojasSanchez_2014_NatComm,Koroteev_2004_PRL,Ast_2007_PRL}, the CPGE of the bilayer may originate from local Cu/Bi interfaces with spin--momentum locked bands of Rashba type. 
(It is not clear how the other planes that exist in the bilayer contribute to the CPGE.)

Previous studies on the CPGE show that the helicity dependent photocurrent is the largest when the light energy is close to the bandgap of the system\cite{Ivchenko_2017_condmat}. In contrast, here we find the CPGE can be induced by light with its energy being much larger than the band gap of Bi. Moreover, the helicity dependent photocurrent is larger for larger light energy.  
It is therefore possible that the origin of the helicity dependent anisotropic photocurrent is different from that of conventional systems, e.g. semiconductor heterostructures. 
We infer the CPGE in Cu/Bi bilayer may originate from spin dependent scattering of the photo-excited carriers via the strong spin orbit coupling of Bi.
The existence of Cu atoms in Bi, which may occur by thermal diffusion, may enhance the spin dependent scattering as Bi doped Cu induces giant spin Hall effect\cite{Niimi_2012_PRL}.

In summary, we have studied the circular photogalvanic effect (CPGE) in Cu/Bi bilayers excited by visible light. We find helicity dependent photocurrent that appears only along the direction perpendicular to the light plane of incidence. 
These results indicate that the spin selective photo-excited carriers in the Cu/Bi bilayer undergo processes that break the symmetry within the film plane and generate the anisotropic photocurrent.
The anisotropy of the helicity dependent photocurrent agrees with the existence of spin--momentum locked bands of Rashba type.
Reference films consisting of single layer of Cu or Bi show reduced signal of the CPGE, suggesting that the spin--momentum locked bands of Rashba type emerge at the Cu/Bi interface.
In contrast to previous studies, however, the CPGE is observed with light excitation energy that is orders of magnitude larger than the characteristic energy of the system, i.e. here the bandgap of Bi.
Furthermore, the helicity dependent anisotropic photocurrent is larger for excitation light with larger energy. 
Although the CPGE of the bilayer may as well originate from spin--momentum locked bands formed at local Cu/Bi interfaces, we infer that it includes contribution from other effects.
To reveal the exact mechanism of the excitation and relaxation processes, further experiments, including studies on the effect of layer thickness, film stacking and magnetic field on the CPGE, are required.\\

\section*{Appendix}
\subsection{Structural characterization}
\begin{figure}[t!]
\includegraphics[scale=0.6]{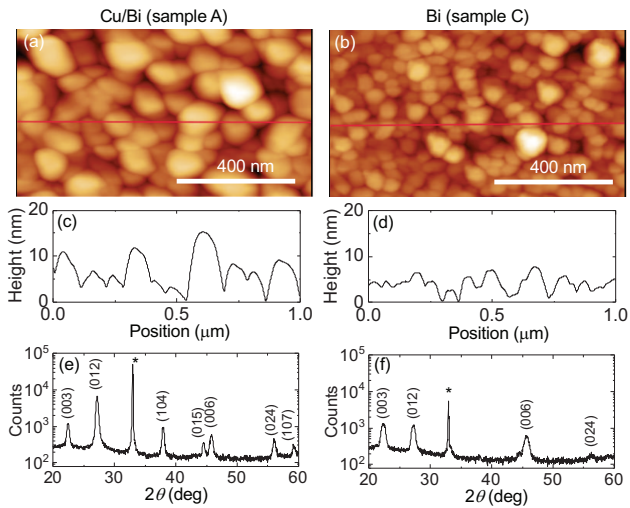}
\caption{(a-f) Atomic force microscopy (AFM) image (a,b), the line profile of the corresponding AFM image along the red line (c,d) and the X-ray diffraction (XRD) spectrum (e,f) of the Cu/Bi bilayer, sample A, (a,c,e) and the Bi single layer, sample C (b,d,f). The indices in (e,f) represent peaks associated with the hexagonal unit cell of Bi. The peak marked with * is due to diffraction from the substrate, i.e. from the Si (002) planes. }
\label{fig:XRD_AFM}
\end{figure}

The atomic force microscopy (AFM) images of the Cu/Bi bilayer (sample A) and the Bi single layer film (sample C) are shown in Figs.~\ref{fig:XRD_AFM}(a) and \ref{fig:XRD_AFM}(b), respectively. Representative line profiles of the corresponding image are shown in Figs.~\ref{fig:XRD_AFM}(c) and \ref{fig:XRD_AFM}(d). The image and the line profile show that grains with lateral diameter of the order $\sim$ 100 nm are formed for the bilayer whereas smaller grains (diameter of $\sim$50 nm) are found for the Bi single layer film. 
As Bi is known to grow in three dimensional form\cite{Missana_1996_APA}, we consider the grains consist of Bi.

The X-ray diffraction (XRD) spectrum of the bilayer and the Bi single layer are shown in Figs.~\ref{fig:XRD_AFM}(e) and \ref{fig:XRD_AFM}(f), respectively. The diffraction peaks are indexed using the hexagonal phase of Bi. The peaks found in the bilayer [Figs.~\ref{fig:XRD_AFM}(e)] correspond to those often found in powder samples, suggesting that the grains are textured but oriented in random directions. Note that the (003) and (006) peaks correspond to diffraction from the plane perpendicular to the c-axis. This plane is known to give relatively large Rashba coupling at surfaces or interfaces between Bi(001) and various materials\cite{Koroteev_2004_PRL,Ast_2007_PRL}.
The XRD spectrum of the Bi single layer film [Figs.~\ref{fig:XRD_AFM}(f)] shows peaks from the c-plane, i.e. the (003) and (006) peaks, and the (012) plane.
The other peaks found in the bilayer are nearly absent. 

It is thus possible that local Cu/Bi(001) interfaces forming spin--momentum locked bands of Rashba type give rise to the helicity dependent anisotropic photocurrent.
With regard to the structural difference of the bilayer and the Bi single layer films, the diffraction peak from the (012) plane is significantly larger for the former.
In addition, diffraction from (104), (015) and (107) planes are only present for the bilayer.
The Bi surface states associated with these planes may possibly contribute to the CPGE of the bilayer.

\begin{acknowledgments}
We thank R. Akiyama for helpful discussions, J. Uzuhashi and T. Ohkubo for technical support. This work was partly supported by JSPS Grant-in-Aid for Scientific Research (16H03853), Specially Promoted Research (15H05702) and the Center of Spintronics Research Network of Japan. Y.-C.L. is supported by JSPS International Fellowship for Research in Japan (Grant No. JP17F17064).
\end{acknowledgments}

\bibliography{ref_102718}

\begin{thebibliography}{38}%
\makeatletter
\providecommand \@ifxundefined [1]{%
 \@ifx{#1\undefined}
}%
\providecommand \@ifnum [1]{%
 \ifnum #1\expandafter \@firstoftwo
 \else \expandafter \@secondoftwo
 \fi
}%
\providecommand \@ifx [1]{%
 \ifx #1\expandafter \@firstoftwo
 \else \expandafter \@secondoftwo
 \fi
}%
\providecommand \natexlab [1]{#1}%
\providecommand \enquote  [1]{``#1''}%
\providecommand \bibnamefont  [1]{#1}%
\providecommand \bibfnamefont [1]{#1}%
\providecommand \citenamefont [1]{#1}%
\providecommand \href@noop [0]{\@secondoftwo}%
\providecommand \href [0]{\begingroup \@sanitize@url \@href}%
\providecommand \@href[1]{\@@startlink{#1}\@@href}%
\providecommand \@@href[1]{\endgroup#1\@@endlink}%
\providecommand \@sanitize@url [0]{\catcode `\\12\catcode `\$12\catcode
  `\&12\catcode `\#12\catcode `\^12\catcode `\_12\catcode `\%12\relax}%
\providecommand \@@startlink[1]{}%
\providecommand \@@endlink[0]{}%
\providecommand \url  [0]{\begingroup\@sanitize@url \@url }%
\providecommand \@url [1]{\endgroup\@href {#1}{\urlprefix }}%
\providecommand \urlprefix  [0]{URL }%
\providecommand \Eprint [0]{\href }%
\providecommand \doibase [0]{http://dx.doi.org/}%
\providecommand \selectlanguage [0]{\@gobble}%
\providecommand \bibinfo  [0]{\@secondoftwo}%
\providecommand \bibfield  [0]{\@secondoftwo}%
\providecommand \translation [1]{[#1]}%
\providecommand \BibitemOpen [0]{}%
\providecommand \bibitemStop [0]{}%
\providecommand \bibitemNoStop [0]{.\EOS\space}%
\providecommand \EOS [0]{\spacefactor3000\relax}%
\providecommand \BibitemShut  [1]{\csname bibitem#1\endcsname}%
\let\auto@bib@innerbib\@empty
\bibitem [{\citenamefont {Hasan}\ and\ \citenamefont
  {Kane}(2010)}]{Hasan_2010_RMP}%
  \BibitemOpen
  \bibfield  {author} {\bibinfo {author} {\bibfnamefont {M.~Z.}\ \bibnamefont
  {Hasan}}\ and\ \bibinfo {author} {\bibfnamefont {C.~L.}\ \bibnamefont
  {Kane}},\ }\href@noop {} {\bibfield  {journal} {\bibinfo  {journal} {Reviews
  of Modern Physics}\ }\textbf {\bibinfo {volume} {82}},\ \bibinfo {pages}
  {3045} (\bibinfo {year} {2010})}\BibitemShut {NoStop}%
\bibitem [{\citenamefont {Sanchez}\ \emph {et~al.}(2013)\citenamefont
  {Sanchez}, \citenamefont {Vila}, \citenamefont {Desfonds}, \citenamefont
  {Gambarelli}, \citenamefont {Attane}, \citenamefont {Teresa}, \citenamefont
  {Magen},\ and\ \citenamefont {Fert}}]{RojasSanchez_2014_NatComm}%
  \BibitemOpen
  \bibfield  {author} {\bibinfo {author} {\bibfnamefont {J.~C.~R.}\
  \bibnamefont {Sanchez}}, \bibinfo {author} {\bibfnamefont {L.}~\bibnamefont
  {Vila}}, \bibinfo {author} {\bibfnamefont {G.}~\bibnamefont {Desfonds}},
  \bibinfo {author} {\bibfnamefont {S.}~\bibnamefont {Gambarelli}}, \bibinfo
  {author} {\bibfnamefont {J.~P.}\ \bibnamefont {Attane}}, \bibinfo {author}
  {\bibfnamefont {J.~M.~D.}\ \bibnamefont {Teresa}}, \bibinfo {author}
  {\bibfnamefont {C.}~\bibnamefont {Magen}}, \ and\ \bibinfo {author}
  {\bibfnamefont {A.}~\bibnamefont {Fert}},\ }\href@noop {} {\bibfield
  {journal} {\bibinfo  {journal} {Nature Communications}\ }\textbf {\bibinfo
  {volume} {4}},\ \bibinfo {pages} {2944} (\bibinfo {year} {2013})}\BibitemShut
  {NoStop}%
\bibitem [{\citenamefont {Lesne}\ \emph {et~al.}(2016)\citenamefont {Lesne},
  \citenamefont {Fu}, \citenamefont {Oyarzun}, \citenamefont {Rojas-Sanchez},
  \citenamefont {Vaz}, \citenamefont {Naganuma}, \citenamefont {Sicoli},
  \citenamefont {Attane}, \citenamefont {Jamet}, \citenamefont {Jacquet},
  \citenamefont {George}, \citenamefont {Barthelemy}, \citenamefont {Jaffres},
  \citenamefont {Fert}, \citenamefont {Bibes},\ and\ \citenamefont
  {Vila}}]{Lesne_2016_NatMater}%
  \BibitemOpen
  \bibfield  {author} {\bibinfo {author} {\bibfnamefont {E.}~\bibnamefont
  {Lesne}}, \bibinfo {author} {\bibfnamefont {Y.}~\bibnamefont {Fu}}, \bibinfo
  {author} {\bibfnamefont {S.}~\bibnamefont {Oyarzun}}, \bibinfo {author}
  {\bibfnamefont {J.~C.}\ \bibnamefont {Rojas-Sanchez}}, \bibinfo {author}
  {\bibfnamefont {D.~C.}\ \bibnamefont {Vaz}}, \bibinfo {author} {\bibfnamefont
  {H.}~\bibnamefont {Naganuma}}, \bibinfo {author} {\bibfnamefont
  {G.}~\bibnamefont {Sicoli}}, \bibinfo {author} {\bibfnamefont {J.~P.}\
  \bibnamefont {Attane}}, \bibinfo {author} {\bibfnamefont {M.}~\bibnamefont
  {Jamet}}, \bibinfo {author} {\bibfnamefont {E.}~\bibnamefont {Jacquet}},
  \bibinfo {author} {\bibfnamefont {J.~M.}\ \bibnamefont {George}}, \bibinfo
  {author} {\bibfnamefont {A.}~\bibnamefont {Barthelemy}}, \bibinfo {author}
  {\bibfnamefont {H.}~\bibnamefont {Jaffres}}, \bibinfo {author} {\bibfnamefont
  {A.}~\bibnamefont {Fert}}, \bibinfo {author} {\bibfnamefont {M.}~\bibnamefont
  {Bibes}}, \ and\ \bibinfo {author} {\bibfnamefont {L.}~\bibnamefont {Vila}},\
  }\href@noop {} {\bibfield  {journal} {\bibinfo  {journal} {Nature Materials}\
  }\textbf {\bibinfo {volume} {15}},\ \bibinfo {pages} {1261} (\bibinfo {year}
  {2016})}\BibitemShut {NoStop}%
\bibitem [{\citenamefont {Song}\ \emph {et~al.}(2017)\citenamefont {Song},
  \citenamefont {Zhang}, \citenamefont {Su}, \citenamefont {Yuan},
  \citenamefont {Chen}, \citenamefont {Xing}, \citenamefont {Shi},
  \citenamefont {Sun},\ and\ \citenamefont {Han}}]{Song_2017_SciAdv}%
  \BibitemOpen
  \bibfield  {author} {\bibinfo {author} {\bibfnamefont {Q.}~\bibnamefont
  {Song}}, \bibinfo {author} {\bibfnamefont {H.~R.}\ \bibnamefont {Zhang}},
  \bibinfo {author} {\bibfnamefont {T.}~\bibnamefont {Su}}, \bibinfo {author}
  {\bibfnamefont {W.}~\bibnamefont {Yuan}}, \bibinfo {author} {\bibfnamefont
  {Y.~Y.}\ \bibnamefont {Chen}}, \bibinfo {author} {\bibfnamefont {W.~Y.}\
  \bibnamefont {Xing}}, \bibinfo {author} {\bibfnamefont {J.}~\bibnamefont
  {Shi}}, \bibinfo {author} {\bibfnamefont {J.~R.}\ \bibnamefont {Sun}}, \ and\
  \bibinfo {author} {\bibfnamefont {W.}~\bibnamefont {Han}},\ }\href@noop {}
  {\bibfield  {journal} {\bibinfo  {journal} {Science Advances}\ }\textbf
  {\bibinfo {volume} {3}},\ \bibinfo {pages} {e1602312} (\bibinfo {year}
  {2017})}\BibitemShut {NoStop}%
\bibitem [{\citenamefont {Bychkov}\ and\ \citenamefont
  {Rashba}(1984)}]{Bychkov_1984_JETP}%
  \BibitemOpen
  \bibfield  {author} {\bibinfo {author} {\bibfnamefont {Y.~A.}\ \bibnamefont
  {Bychkov}}\ and\ \bibinfo {author} {\bibfnamefont {E.~I.}\ \bibnamefont
  {Rashba}},\ }\href@noop {} {\bibfield  {journal} {\bibinfo  {journal} {Jetp
  Lett.}\ }\textbf {\bibinfo {volume} {39}},\ \bibinfo {pages} {78} (\bibinfo
  {year} {1984})}\BibitemShut {NoStop}%
\bibitem [{\citenamefont {Manchon}\ \emph {et~al.}(2015)\citenamefont
  {Manchon}, \citenamefont {Koo}, \citenamefont {Nitta}, \citenamefont
  {Frolov},\ and\ \citenamefont {Duine}}]{Manchon_2015_NatMater}%
  \BibitemOpen
  \bibfield  {author} {\bibinfo {author} {\bibfnamefont {A.}~\bibnamefont
  {Manchon}}, \bibinfo {author} {\bibfnamefont {H.~C.}\ \bibnamefont {Koo}},
  \bibinfo {author} {\bibfnamefont {J.}~\bibnamefont {Nitta}}, \bibinfo
  {author} {\bibfnamefont {S.~M.}\ \bibnamefont {Frolov}}, \ and\ \bibinfo
  {author} {\bibfnamefont {R.~A.}\ \bibnamefont {Duine}},\ }\href@noop {}
  {\bibfield  {journal} {\bibinfo  {journal} {Nat. Mater.}\ }\textbf {\bibinfo
  {volume} {14}},\ \bibinfo {pages} {871} (\bibinfo {year} {2015})}\BibitemShut
  {NoStop}%
\bibitem [{\citenamefont {Edelstein}(1990)}]{Edelstein_1990_SSC}%
  \BibitemOpen
  \bibfield  {author} {\bibinfo {author} {\bibfnamefont {V.~M.}\ \bibnamefont
  {Edelstein}},\ }\href@noop {} {\bibfield  {journal} {\bibinfo  {journal}
  {Solid State Commun.}\ }\textbf {\bibinfo {volume} {73}},\ \bibinfo {pages}
  {233} (\bibinfo {year} {1990})}\BibitemShut {NoStop}%
\bibitem [{\citenamefont {Kato}\ \emph {et~al.}(2004)\citenamefont {Kato},
  \citenamefont {Myers}, \citenamefont {Gossard},\ and\ \citenamefont
  {Awschalom}}]{Kato_2004_PRL}%
  \BibitemOpen
  \bibfield  {author} {\bibinfo {author} {\bibfnamefont {Y.~K.}\ \bibnamefont
  {Kato}}, \bibinfo {author} {\bibfnamefont {R.~C.}\ \bibnamefont {Myers}},
  \bibinfo {author} {\bibfnamefont {A.~C.}\ \bibnamefont {Gossard}}, \ and\
  \bibinfo {author} {\bibfnamefont {D.~D.}\ \bibnamefont {Awschalom}},\
  }\href@noop {} {\bibfield  {journal} {\bibinfo  {journal} {Phys. Rev. Lett.}\
  }\textbf {\bibinfo {volume} {93}},\ \bibinfo {pages} {176601} (\bibinfo
  {year} {2004})}\BibitemShut {NoStop}%
\bibitem [{\citenamefont {Nitta}\ \emph {et~al.}(1997)\citenamefont {Nitta},
  \citenamefont {Akazaki}, \citenamefont {Takayanagi},\ and\ \citenamefont
  {Enoki}}]{Nitta_1997_PRL}%
  \BibitemOpen
  \bibfield  {author} {\bibinfo {author} {\bibfnamefont {J.}~\bibnamefont
  {Nitta}}, \bibinfo {author} {\bibfnamefont {T.}~\bibnamefont {Akazaki}},
  \bibinfo {author} {\bibfnamefont {H.}~\bibnamefont {Takayanagi}}, \ and\
  \bibinfo {author} {\bibfnamefont {T.}~\bibnamefont {Enoki}},\ }\href@noop {}
  {\bibfield  {journal} {\bibinfo  {journal} {Physical Review Letters}\
  }\textbf {\bibinfo {volume} {78}},\ \bibinfo {pages} {1335} (\bibinfo {year}
  {1997})}\BibitemShut {NoStop}%
\bibitem [{\citenamefont {Zutic}\ \emph {et~al.}(2004)\citenamefont {Zutic},
  \citenamefont {Fabian},\ and\ \citenamefont {Sarma}}]{Zutic_2004_RMP}%
  \BibitemOpen
  \bibfield  {author} {\bibinfo {author} {\bibfnamefont {I.}~\bibnamefont
  {Zutic}}, \bibinfo {author} {\bibfnamefont {J.}~\bibnamefont {Fabian}}, \
  and\ \bibinfo {author} {\bibfnamefont {S.~D.}\ \bibnamefont {Sarma}},\
  }\href@noop {} {\bibfield  {journal} {\bibinfo  {journal} {Rev. Mod. Phys.}\
  }\textbf {\bibinfo {volume} {76}},\ \bibinfo {pages} {323} (\bibinfo {year}
  {2004})}\BibitemShut {NoStop}%
\bibitem [{\citenamefont {Isasa}\ \emph {et~al.}(2016)\citenamefont {Isasa},
  \citenamefont {Martinez-Velarte}, \citenamefont {Villamor}, \citenamefont
  {Magen}, \citenamefont {Morellon}, \citenamefont {Teresa}, \citenamefont
  {Ibarra}, \citenamefont {Vignale}, \citenamefont {Chulkov}, \citenamefont
  {Krasovskii}, \citenamefont {Hueso},\ and\ \citenamefont
  {Casanova}}]{Isasa_2016_PRB}%
  \BibitemOpen
  \bibfield  {author} {\bibinfo {author} {\bibfnamefont {M.}~\bibnamefont
  {Isasa}}, \bibinfo {author} {\bibfnamefont {M.~C.}\ \bibnamefont
  {Martinez-Velarte}}, \bibinfo {author} {\bibfnamefont {E.}~\bibnamefont
  {Villamor}}, \bibinfo {author} {\bibfnamefont {C.}~\bibnamefont {Magen}},
  \bibinfo {author} {\bibfnamefont {L.}~\bibnamefont {Morellon}}, \bibinfo
  {author} {\bibfnamefont {J.~M.~D.}\ \bibnamefont {Teresa}}, \bibinfo {author}
  {\bibfnamefont {M.~R.}\ \bibnamefont {Ibarra}}, \bibinfo {author}
  {\bibfnamefont {G.}~\bibnamefont {Vignale}}, \bibinfo {author} {\bibfnamefont
  {E.~V.}\ \bibnamefont {Chulkov}}, \bibinfo {author} {\bibfnamefont {E.~E.}\
  \bibnamefont {Krasovskii}}, \bibinfo {author} {\bibfnamefont {L.~E.}\
  \bibnamefont {Hueso}}, \ and\ \bibinfo {author} {\bibfnamefont
  {F.}~\bibnamefont {Casanova}},\ }\href@noop {} {\bibfield  {journal}
  {\bibinfo  {journal} {Physical Review B}\ }\textbf {\bibinfo {volume} {93}},\
  \bibinfo {pages} {014420} (\bibinfo {year} {2016})}\BibitemShut {NoStop}%
\bibitem [{\citenamefont {Matsushima}\ \emph {et~al.}(2017)\citenamefont
  {Matsushima}, \citenamefont {Ando}, \citenamefont {Dushenko}, \citenamefont
  {Ohshima}, \citenamefont {Kumamoto}, \citenamefont {Shinjo},\ and\
  \citenamefont {Shiraishi}}]{Matsushima_2017_APL}%
  \BibitemOpen
  \bibfield  {author} {\bibinfo {author} {\bibfnamefont {M.}~\bibnamefont
  {Matsushima}}, \bibinfo {author} {\bibfnamefont {Y.}~\bibnamefont {Ando}},
  \bibinfo {author} {\bibfnamefont {S.}~\bibnamefont {Dushenko}}, \bibinfo
  {author} {\bibfnamefont {R.}~\bibnamefont {Ohshima}}, \bibinfo {author}
  {\bibfnamefont {R.}~\bibnamefont {Kumamoto}}, \bibinfo {author}
  {\bibfnamefont {T.}~\bibnamefont {Shinjo}}, \ and\ \bibinfo {author}
  {\bibfnamefont {M.}~\bibnamefont {Shiraishi}},\ }\href@noop {} {\bibfield
  {journal} {\bibinfo  {journal} {Applied Physics Letters}\ }\textbf {\bibinfo
  {volume} {110}},\ \bibinfo {pages} {072404} (\bibinfo {year}
  {2017})}\BibitemShut {NoStop}%
\bibitem [{\citenamefont {Miron}\ \emph
  {et~al.}(2011{\natexlab{a}})\citenamefont {Miron}, \citenamefont {Moore},
  \citenamefont {Szambolics}, \citenamefont {Buda-Prejbeanu}, \citenamefont
  {Auffret}, \citenamefont {Rodmacq}, \citenamefont {Pizzini}, \citenamefont
  {Vogel}, \citenamefont {Bonfim}, \citenamefont {Schuhl},\ and\ \citenamefont
  {Gaudin}}]{Miron_2011_NatMater}%
  \BibitemOpen
  \bibfield  {author} {\bibinfo {author} {\bibfnamefont {I.~M.}\ \bibnamefont
  {Miron}}, \bibinfo {author} {\bibfnamefont {T.}~\bibnamefont {Moore}},
  \bibinfo {author} {\bibfnamefont {H.}~\bibnamefont {Szambolics}}, \bibinfo
  {author} {\bibfnamefont {L.~D.}\ \bibnamefont {Buda-Prejbeanu}}, \bibinfo
  {author} {\bibfnamefont {S.}~\bibnamefont {Auffret}}, \bibinfo {author}
  {\bibfnamefont {B.}~\bibnamefont {Rodmacq}}, \bibinfo {author} {\bibfnamefont
  {S.}~\bibnamefont {Pizzini}}, \bibinfo {author} {\bibfnamefont
  {J.}~\bibnamefont {Vogel}}, \bibinfo {author} {\bibfnamefont
  {M.}~\bibnamefont {Bonfim}}, \bibinfo {author} {\bibfnamefont
  {A.}~\bibnamefont {Schuhl}}, \ and\ \bibinfo {author} {\bibfnamefont
  {G.}~\bibnamefont {Gaudin}},\ }\href@noop {} {\bibfield  {journal} {\bibinfo
  {journal} {Nat. Mater.}\ }\textbf {\bibinfo {volume} {10}},\ \bibinfo {pages}
  {419} (\bibinfo {year} {2011}{\natexlab{a}})}\BibitemShut {NoStop}%
\bibitem [{\citenamefont {Miron}\ \emph
  {et~al.}(2011{\natexlab{b}})\citenamefont {Miron}, \citenamefont {Garello},
  \citenamefont {Gaudin}, \citenamefont {Zermatten}, \citenamefont {Costache},
  \citenamefont {Auffret}, \citenamefont {Bandiera}, \citenamefont {Rodmacq},
  \citenamefont {Schuhl},\ and\ \citenamefont
  {Gambardella}}]{Miron_2011_Nature}%
  \BibitemOpen
  \bibfield  {author} {\bibinfo {author} {\bibfnamefont {I.~M.}\ \bibnamefont
  {Miron}}, \bibinfo {author} {\bibfnamefont {K.}~\bibnamefont {Garello}},
  \bibinfo {author} {\bibfnamefont {G.}~\bibnamefont {Gaudin}}, \bibinfo
  {author} {\bibfnamefont {P.~J.}\ \bibnamefont {Zermatten}}, \bibinfo {author}
  {\bibfnamefont {M.~V.}\ \bibnamefont {Costache}}, \bibinfo {author}
  {\bibfnamefont {S.}~\bibnamefont {Auffret}}, \bibinfo {author} {\bibfnamefont
  {S.}~\bibnamefont {Bandiera}}, \bibinfo {author} {\bibfnamefont
  {B.}~\bibnamefont {Rodmacq}}, \bibinfo {author} {\bibfnamefont
  {A.}~\bibnamefont {Schuhl}}, \ and\ \bibinfo {author} {\bibfnamefont
  {P.}~\bibnamefont {Gambardella}},\ }\href@noop {} {\bibfield  {journal}
  {\bibinfo  {journal} {Nature}\ }\textbf {\bibinfo {volume} {476}},\ \bibinfo
  {pages} {189} (\bibinfo {year} {2011}{\natexlab{b}})}\BibitemShut {NoStop}%
\bibitem [{\citenamefont {Ganichev}\ \emph {et~al.}(2002)\citenamefont
  {Ganichev}, \citenamefont {Ivchenko}, \citenamefont {Bel'kov}, \citenamefont
  {Tarasenko}, \citenamefont {Sollinger}, \citenamefont {Weiss}, \citenamefont
  {Wegscheider},\ and\ \citenamefont {Prettl}}]{Ganichev_2002_Nature}%
  \BibitemOpen
  \bibfield  {author} {\bibinfo {author} {\bibfnamefont {S.~D.}\ \bibnamefont
  {Ganichev}}, \bibinfo {author} {\bibfnamefont {E.~L.}\ \bibnamefont
  {Ivchenko}}, \bibinfo {author} {\bibfnamefont {V.~V.}\ \bibnamefont
  {Bel'kov}}, \bibinfo {author} {\bibfnamefont {S.~A.}\ \bibnamefont
  {Tarasenko}}, \bibinfo {author} {\bibfnamefont {M.}~\bibnamefont
  {Sollinger}}, \bibinfo {author} {\bibfnamefont {D.}~\bibnamefont {Weiss}},
  \bibinfo {author} {\bibfnamefont {W.}~\bibnamefont {Wegscheider}}, \ and\
  \bibinfo {author} {\bibfnamefont {W.}~\bibnamefont {Prettl}},\ }\href@noop {}
  {\bibfield  {journal} {\bibinfo  {journal} {Nature}\ }\textbf {\bibinfo
  {volume} {417}},\ \bibinfo {pages} {153} (\bibinfo {year}
  {2002})}\BibitemShut {NoStop}%
\bibitem [{\citenamefont {Ivchenko}\ and\ \citenamefont
  {Ganichev}()}]{Ivchenko_2017_condmat}%
  \BibitemOpen
  \bibfield  {author} {\bibinfo {author} {\bibfnamefont {E.}~\bibnamefont
  {Ivchenko}}\ and\ \bibinfo {author} {\bibfnamefont {S.}~\bibnamefont
  {Ganichev}},\ }\href@noop {} {\ ,\ \bibinfo {pages}
  {arXiv:1710.09223}}\BibitemShut {NoStop}%
\bibitem [{\citenamefont {Ganichev}\ \emph {et~al.}(2004)\citenamefont
  {Ganichev}, \citenamefont {Bel'kov}, \citenamefont {Golub}, \citenamefont
  {Ivchenko}, \citenamefont {Schneider}, \citenamefont {Giglberger},
  \citenamefont {Eroms}, \citenamefont {Boeck}, \citenamefont {Borghs},
  \citenamefont {Wegscheider}, \citenamefont {Weiss},\ and\ \citenamefont
  {Prettl}}]{Ganichev_2004_PRL}%
  \BibitemOpen
  \bibfield  {author} {\bibinfo {author} {\bibfnamefont {S.~D.}\ \bibnamefont
  {Ganichev}}, \bibinfo {author} {\bibfnamefont {V.~V.}\ \bibnamefont
  {Bel'kov}}, \bibinfo {author} {\bibfnamefont {L.~E.}\ \bibnamefont {Golub}},
  \bibinfo {author} {\bibfnamefont {E.~L.}\ \bibnamefont {Ivchenko}}, \bibinfo
  {author} {\bibfnamefont {P.}~\bibnamefont {Schneider}}, \bibinfo {author}
  {\bibfnamefont {S.}~\bibnamefont {Giglberger}}, \bibinfo {author}
  {\bibfnamefont {J.}~\bibnamefont {Eroms}}, \bibinfo {author} {\bibfnamefont
  {J.~D.}\ \bibnamefont {Boeck}}, \bibinfo {author} {\bibfnamefont
  {G.}~\bibnamefont {Borghs}}, \bibinfo {author} {\bibfnamefont
  {W.}~\bibnamefont {Wegscheider}}, \bibinfo {author} {\bibfnamefont
  {D.}~\bibnamefont {Weiss}}, \ and\ \bibinfo {author} {\bibfnamefont
  {W.}~\bibnamefont {Prettl}},\ }\href@noop {} {\bibfield  {journal} {\bibinfo
  {journal} {Physical Review Letters}\ }\textbf {\bibinfo {volume} {92}},\
  \bibinfo {pages} {256601} (\bibinfo {year} {2004})}\BibitemShut {NoStop}%
\bibitem [{\citenamefont {Weber}\ \emph {et~al.}(2005)\citenamefont {Weber},
  \citenamefont {Ganichev}, \citenamefont {Danilov}, \citenamefont {Weiss},
  \citenamefont {Prettl}, \citenamefont {Kvon}, \citenamefont {Bel'kov},
  \citenamefont {Golub}, \citenamefont {Cho},\ and\ \citenamefont
  {Lee}}]{Weber_2005_APL}%
  \BibitemOpen
  \bibfield  {author} {\bibinfo {author} {\bibfnamefont {W.}~\bibnamefont
  {Weber}}, \bibinfo {author} {\bibfnamefont {S.~D.}\ \bibnamefont {Ganichev}},
  \bibinfo {author} {\bibfnamefont {S.~N.}\ \bibnamefont {Danilov}}, \bibinfo
  {author} {\bibfnamefont {D.}~\bibnamefont {Weiss}}, \bibinfo {author}
  {\bibfnamefont {W.}~\bibnamefont {Prettl}}, \bibinfo {author} {\bibfnamefont
  {Z.~D.}\ \bibnamefont {Kvon}}, \bibinfo {author} {\bibfnamefont {V.~V.}\
  \bibnamefont {Bel'kov}}, \bibinfo {author} {\bibfnamefont {L.~E.}\
  \bibnamefont {Golub}}, \bibinfo {author} {\bibfnamefont {H.~I.}\ \bibnamefont
  {Cho}}, \ and\ \bibinfo {author} {\bibfnamefont {J.~H.}\ \bibnamefont
  {Lee}},\ }\href@noop {} {\bibfield  {journal} {\bibinfo  {journal} {Applied
  Physics Letters}\ }\textbf {\bibinfo {volume} {87}},\ \bibinfo {pages}
  {262106} (\bibinfo {year} {2005})}\BibitemShut {NoStop}%
\bibitem [{\citenamefont {Kohda}\ \emph {et~al.}(2015)\citenamefont {Kohda},
  \citenamefont {Altmann}, \citenamefont {Schuh}, \citenamefont {Ganichev},
  \citenamefont {Wegscheider},\ and\ \citenamefont {Salis}}]{Kohda_2015_APL}%
  \BibitemOpen
  \bibfield  {author} {\bibinfo {author} {\bibfnamefont {M.}~\bibnamefont
  {Kohda}}, \bibinfo {author} {\bibfnamefont {P.}~\bibnamefont {Altmann}},
  \bibinfo {author} {\bibfnamefont {D.}~\bibnamefont {Schuh}}, \bibinfo
  {author} {\bibfnamefont {S.~D.}\ \bibnamefont {Ganichev}}, \bibinfo {author}
  {\bibfnamefont {W.}~\bibnamefont {Wegscheider}}, \ and\ \bibinfo {author}
  {\bibfnamefont {G.}~\bibnamefont {Salis}},\ }\href@noop {} {\bibfield
  {journal} {\bibinfo  {journal} {Applied Physics Letters}\ }\textbf {\bibinfo
  {volume} {107}},\ \bibinfo {pages} {172402} (\bibinfo {year}
  {2015})}\BibitemShut {NoStop}%
\bibitem [{\citenamefont {Hosur}(2011)}]{Hosur_2011_PRB}%
  \BibitemOpen
  \bibfield  {author} {\bibinfo {author} {\bibfnamefont {P.}~\bibnamefont
  {Hosur}},\ }\href@noop {} {\bibfield  {journal} {\bibinfo  {journal}
  {Physical Review B}\ }\textbf {\bibinfo {volume} {83}},\ \bibinfo {pages}
  {035309} (\bibinfo {year} {2011})}\BibitemShut {NoStop}%
\bibitem [{\citenamefont {Kastl}\ \emph {et~al.}(2015)\citenamefont {Kastl},
  \citenamefont {Karnetzky}, \citenamefont {Karl},\ and\ \citenamefont
  {Holleitner}}]{Kastl_2015_NatComm}%
  \BibitemOpen
  \bibfield  {author} {\bibinfo {author} {\bibfnamefont {C.}~\bibnamefont
  {Kastl}}, \bibinfo {author} {\bibfnamefont {C.}~\bibnamefont {Karnetzky}},
  \bibinfo {author} {\bibfnamefont {H.}~\bibnamefont {Karl}}, \ and\ \bibinfo
  {author} {\bibfnamefont {A.~W.}\ \bibnamefont {Holleitner}},\ }\href@noop {}
  {\bibfield  {journal} {\bibinfo  {journal} {Nature Communications}\ }\textbf
  {\bibinfo {volume} {6}},\ \bibinfo {pages} {6617} (\bibinfo {year}
  {2015})}\BibitemShut {NoStop}%
\bibitem [{\citenamefont {Okada}\ \emph {et~al.}(2016)\citenamefont {Okada},
  \citenamefont {Ogawa}, \citenamefont {Yoshimi}, \citenamefont {Tsukazaki},
  \citenamefont {Takahashi}, \citenamefont {Kawasaki},\ and\ \citenamefont
  {Tokura}}]{Okada_2016_PRB}%
  \BibitemOpen
  \bibfield  {author} {\bibinfo {author} {\bibfnamefont {K.~N.}\ \bibnamefont
  {Okada}}, \bibinfo {author} {\bibfnamefont {N.}~\bibnamefont {Ogawa}},
  \bibinfo {author} {\bibfnamefont {R.}~\bibnamefont {Yoshimi}}, \bibinfo
  {author} {\bibfnamefont {A.}~\bibnamefont {Tsukazaki}}, \bibinfo {author}
  {\bibfnamefont {K.~S.}\ \bibnamefont {Takahashi}}, \bibinfo {author}
  {\bibfnamefont {M.}~\bibnamefont {Kawasaki}}, \ and\ \bibinfo {author}
  {\bibfnamefont {Y.}~\bibnamefont {Tokura}},\ }\href@noop {} {\bibfield
  {journal} {\bibinfo  {journal} {Physical Review B}\ }\textbf {\bibinfo
  {volume} {93}},\ \bibinfo {pages} {081403} (\bibinfo {year}
  {2016})}\BibitemShut {NoStop}%
\bibitem [{\citenamefont {Pan}\ \emph {et~al.}(2017)\citenamefont {Pan},
  \citenamefont {Wang}, \citenamefont {Yeats}, \citenamefont {Pillsbury},
  \citenamefont {Flanagan}, \citenamefont {Richardella}, \citenamefont {Zhang},
  \citenamefont {Awschalom}, \citenamefont {Liu},\ and\ \citenamefont
  {Samarth}}]{Pan_2017_NatComm}%
  \BibitemOpen
  \bibfield  {author} {\bibinfo {author} {\bibfnamefont {Y.}~\bibnamefont
  {Pan}}, \bibinfo {author} {\bibfnamefont {Q.~Z.}\ \bibnamefont {Wang}},
  \bibinfo {author} {\bibfnamefont {A.~L.}\ \bibnamefont {Yeats}}, \bibinfo
  {author} {\bibfnamefont {T.}~\bibnamefont {Pillsbury}}, \bibinfo {author}
  {\bibfnamefont {T.~C.}\ \bibnamefont {Flanagan}}, \bibinfo {author}
  {\bibfnamefont {A.}~\bibnamefont {Richardella}}, \bibinfo {author}
  {\bibfnamefont {H.~J.}\ \bibnamefont {Zhang}}, \bibinfo {author}
  {\bibfnamefont {D.~D.}\ \bibnamefont {Awschalom}}, \bibinfo {author}
  {\bibfnamefont {C.~X.}\ \bibnamefont {Liu}}, \ and\ \bibinfo {author}
  {\bibfnamefont {N.}~\bibnamefont {Samarth}},\ }\href@noop {} {\bibfield
  {journal} {\bibinfo  {journal} {Nature Communications}\ }\textbf {\bibinfo
  {volume} {8}},\ \bibinfo {pages} {1037} (\bibinfo {year} {2017})}\BibitemShut
  {NoStop}%
\bibitem [{\citenamefont {Ma}\ \emph {et~al.}(2017)\citenamefont {Ma},
  \citenamefont {Xu}, \citenamefont {Chan}, \citenamefont {Zhang},
  \citenamefont {Chang}, \citenamefont {Lin}, \citenamefont {Xie},
  \citenamefont {Palacios}, \citenamefont {Lin}, \citenamefont {Jia},
  \citenamefont {Lee}, \citenamefont {Jarillo-Herrero},\ and\ \citenamefont
  {Gedik}}]{Ma_2017_NatPhys}%
  \BibitemOpen
  \bibfield  {author} {\bibinfo {author} {\bibfnamefont {Q.}~\bibnamefont
  {Ma}}, \bibinfo {author} {\bibfnamefont {S.~Y.}\ \bibnamefont {Xu}}, \bibinfo
  {author} {\bibfnamefont {C.~K.}\ \bibnamefont {Chan}}, \bibinfo {author}
  {\bibfnamefont {C.~L.}\ \bibnamefont {Zhang}}, \bibinfo {author}
  {\bibfnamefont {G.~Q.}\ \bibnamefont {Chang}}, \bibinfo {author}
  {\bibfnamefont {Y.~X.}\ \bibnamefont {Lin}}, \bibinfo {author} {\bibfnamefont
  {W.~W.}\ \bibnamefont {Xie}}, \bibinfo {author} {\bibfnamefont
  {T.}~\bibnamefont {Palacios}}, \bibinfo {author} {\bibfnamefont
  {H.}~\bibnamefont {Lin}}, \bibinfo {author} {\bibfnamefont {S.}~\bibnamefont
  {Jia}}, \bibinfo {author} {\bibfnamefont {P.~A.}\ \bibnamefont {Lee}},
  \bibinfo {author} {\bibfnamefont {P.}~\bibnamefont {Jarillo-Herrero}}, \ and\
  \bibinfo {author} {\bibfnamefont {N.}~\bibnamefont {Gedik}},\ }\href@noop {}
  {\bibfield  {journal} {\bibinfo  {journal} {Nature Physics}\ }\textbf
  {\bibinfo {volume} {13}},\ \bibinfo {pages} {842} (\bibinfo {year}
  {2017})}\BibitemShut {NoStop}%
\bibitem [{\citenamefont {McIver}\ \emph {et~al.}(2012)\citenamefont {McIver},
  \citenamefont {Hsieh}, \citenamefont {Steinberg}, \citenamefont
  {Jarillo-Herrero},\ and\ \citenamefont {Gedik}}]{McIver_2012_NatNano}%
  \BibitemOpen
  \bibfield  {author} {\bibinfo {author} {\bibfnamefont {J.~W.}\ \bibnamefont
  {McIver}}, \bibinfo {author} {\bibfnamefont {D.}~\bibnamefont {Hsieh}},
  \bibinfo {author} {\bibfnamefont {H.}~\bibnamefont {Steinberg}}, \bibinfo
  {author} {\bibfnamefont {P.}~\bibnamefont {Jarillo-Herrero}}, \ and\ \bibinfo
  {author} {\bibfnamefont {N.}~\bibnamefont {Gedik}},\ }\href@noop {}
  {\bibfield  {journal} {\bibinfo  {journal} {Nature Nanotechnology}\ }\textbf
  {\bibinfo {volume} {7}},\ \bibinfo {pages} {96} (\bibinfo {year}
  {2012})}\BibitemShut {NoStop}%
\bibitem [{\citenamefont {Yuan}\ \emph {et~al.}(2014)\citenamefont {Yuan},
  \citenamefont {Wang}, \citenamefont {Lian}, \citenamefont {Zhang},
  \citenamefont {Fang}, \citenamefont {Shen}, \citenamefont {Xu}, \citenamefont
  {Xu}, \citenamefont {Zhang}, \citenamefont {Hwang},\ and\ \citenamefont
  {Cui}}]{Yuan_2014_NatNano}%
  \BibitemOpen
  \bibfield  {author} {\bibinfo {author} {\bibfnamefont {H.~T.}\ \bibnamefont
  {Yuan}}, \bibinfo {author} {\bibfnamefont {X.~Q.}\ \bibnamefont {Wang}},
  \bibinfo {author} {\bibfnamefont {B.}~\bibnamefont {Lian}}, \bibinfo {author}
  {\bibfnamefont {H.~J.}\ \bibnamefont {Zhang}}, \bibinfo {author}
  {\bibfnamefont {X.~F.}\ \bibnamefont {Fang}}, \bibinfo {author}
  {\bibfnamefont {B.}~\bibnamefont {Shen}}, \bibinfo {author} {\bibfnamefont
  {G.}~\bibnamefont {Xu}}, \bibinfo {author} {\bibfnamefont {Y.}~\bibnamefont
  {Xu}}, \bibinfo {author} {\bibfnamefont {S.~C.}\ \bibnamefont {Zhang}},
  \bibinfo {author} {\bibfnamefont {H.~Y.}\ \bibnamefont {Hwang}}, \ and\
  \bibinfo {author} {\bibfnamefont {Y.}~\bibnamefont {Cui}},\ }\href@noop {}
  {\bibfield  {journal} {\bibinfo  {journal} {Nature Nanotechnology}\ }\textbf
  {\bibinfo {volume} {9}},\ \bibinfo {pages} {851} (\bibinfo {year}
  {2014})}\BibitemShut {NoStop}%
\bibitem [{\citenamefont {Zutic}\ \emph {et~al.}(2002)\citenamefont {Zutic},
  \citenamefont {Fabian},\ and\ \citenamefont {Sarma}}]{Zutic_2002_PRL}%
  \BibitemOpen
  \bibfield  {author} {\bibinfo {author} {\bibfnamefont {I.}~\bibnamefont
  {Zutic}}, \bibinfo {author} {\bibfnamefont {J.}~\bibnamefont {Fabian}}, \
  and\ \bibinfo {author} {\bibfnamefont {S.~D.}\ \bibnamefont {Sarma}},\
  }\href@noop {} {\bibfield  {journal} {\bibinfo  {journal} {Physical Review
  Letters}\ }\textbf {\bibinfo {volume} {88}},\ \bibinfo {pages} {066603}
  (\bibinfo {year} {2002})}\BibitemShut {NoStop}%
\bibitem [{\citenamefont {Zutic}\ \emph {et~al.}(2006)\citenamefont {Zutic},
  \citenamefont {Fabian},\ and\ \citenamefont {Erwin}}]{Zutic_2006_PRL}%
  \BibitemOpen
  \bibfield  {author} {\bibinfo {author} {\bibfnamefont {I.}~\bibnamefont
  {Zutic}}, \bibinfo {author} {\bibfnamefont {J.}~\bibnamefont {Fabian}}, \
  and\ \bibinfo {author} {\bibfnamefont {S.~C.}\ \bibnamefont {Erwin}},\
  }\href@noop {} {\bibfield  {journal} {\bibinfo  {journal} {Physical Review
  Letters}\ }\textbf {\bibinfo {volume} {97}},\ \bibinfo {pages} {026602}
  (\bibinfo {year} {2006})}\BibitemShut {NoStop}%
\bibitem [{\citenamefont {Kondo}\ \emph {et~al.}(2006)\citenamefont {Kondo},
  \citenamefont {Hayafuji},\ and\ \citenamefont {Munekata}}]{Kondo_2006_JJAP}%
  \BibitemOpen
  \bibfield  {author} {\bibinfo {author} {\bibfnamefont {T.}~\bibnamefont
  {Kondo}}, \bibinfo {author} {\bibfnamefont {J.~J.}\ \bibnamefont {Hayafuji}},
  \ and\ \bibinfo {author} {\bibfnamefont {H.}~\bibnamefont {Munekata}},\
  }\href@noop {} {\bibfield  {journal} {\bibinfo  {journal} {Japanese Journal
  of Applied Physics}\ }\textbf {\bibinfo {volume} {45}},\ \bibinfo {pages}
  {L663} (\bibinfo {year} {2006})}\BibitemShut {NoStop}%
\bibitem [{\citenamefont {Endres}\ \emph {et~al.}(2013)\citenamefont {Endres},
  \citenamefont {Ciorga}, \citenamefont {Schmid}, \citenamefont {Utz},
  \citenamefont {Bougeard}, \citenamefont {Weiss}, \citenamefont {Bayreuther},\
  and\ \citenamefont {Back}}]{Endres_2013_NatComm}%
  \BibitemOpen
  \bibfield  {author} {\bibinfo {author} {\bibfnamefont {B.}~\bibnamefont
  {Endres}}, \bibinfo {author} {\bibfnamefont {M.}~\bibnamefont {Ciorga}},
  \bibinfo {author} {\bibfnamefont {M.}~\bibnamefont {Schmid}}, \bibinfo
  {author} {\bibfnamefont {M.}~\bibnamefont {Utz}}, \bibinfo {author}
  {\bibfnamefont {D.}~\bibnamefont {Bougeard}}, \bibinfo {author}
  {\bibfnamefont {D.}~\bibnamefont {Weiss}}, \bibinfo {author} {\bibfnamefont
  {G.}~\bibnamefont {Bayreuther}}, \ and\ \bibinfo {author} {\bibfnamefont
  {C.~H.}\ \bibnamefont {Back}},\ }\href@noop {} {\bibfield  {journal}
  {\bibinfo  {journal} {Nature Communications}\ }\textbf {\bibinfo {volume}
  {4}},\ \bibinfo {pages} {2068} (\bibinfo {year} {2013})}\BibitemShut
  {NoStop}%
\bibitem [{\citenamefont {Olbrich}\ \emph {et~al.}(2014)\citenamefont
  {Olbrich}, \citenamefont {Golub}, \citenamefont {Herrmann}, \citenamefont
  {Danilov}, \citenamefont {Plank}, \citenamefont {Bel'kov}, \citenamefont
  {Mussler}, \citenamefont {Weyrich}, \citenamefont {Schneider}, \citenamefont
  {Kampmeier}, \citenamefont {Grutzmacher}, \citenamefont {Plucinski},
  \citenamefont {Eschbach},\ and\ \citenamefont {Ganichev}}]{Olbrich_2014_PRL}%
  \BibitemOpen
  \bibfield  {author} {\bibinfo {author} {\bibfnamefont {P.}~\bibnamefont
  {Olbrich}}, \bibinfo {author} {\bibfnamefont {L.~E.}\ \bibnamefont {Golub}},
  \bibinfo {author} {\bibfnamefont {T.}~\bibnamefont {Herrmann}}, \bibinfo
  {author} {\bibfnamefont {S.~N.}\ \bibnamefont {Danilov}}, \bibinfo {author}
  {\bibfnamefont {H.}~\bibnamefont {Plank}}, \bibinfo {author} {\bibfnamefont
  {V.~V.}\ \bibnamefont {Bel'kov}}, \bibinfo {author} {\bibfnamefont
  {G.}~\bibnamefont {Mussler}}, \bibinfo {author} {\bibfnamefont
  {C.}~\bibnamefont {Weyrich}}, \bibinfo {author} {\bibfnamefont {C.~M.}\
  \bibnamefont {Schneider}}, \bibinfo {author} {\bibfnamefont {J.}~\bibnamefont
  {Kampmeier}}, \bibinfo {author} {\bibfnamefont {D.}~\bibnamefont
  {Grutzmacher}}, \bibinfo {author} {\bibfnamefont {L.}~\bibnamefont
  {Plucinski}}, \bibinfo {author} {\bibfnamefont {M.}~\bibnamefont {Eschbach}},
  \ and\ \bibinfo {author} {\bibfnamefont {S.~D.}\ \bibnamefont {Ganichev}},\
  }\href@noop {} {\bibfield  {journal} {\bibinfo  {journal} {Physical Review
  Letters}\ }\textbf {\bibinfo {volume} {113}},\ \bibinfo {pages} {096601}
  (\bibinfo {year} {2014})}\BibitemShut {NoStop}%
\bibitem [{\citenamefont {Liu}\ and\ \citenamefont
  {Allen}(1995)}]{Liu_1995_PRB}%
  \BibitemOpen
  \bibfield  {author} {\bibinfo {author} {\bibfnamefont {Y.}~\bibnamefont
  {Liu}}\ and\ \bibinfo {author} {\bibfnamefont {R.~E.}\ \bibnamefont
  {Allen}},\ }\href@noop {} {\bibfield  {journal} {\bibinfo  {journal}
  {Physical Review B}\ }\textbf {\bibinfo {volume} {52}},\ \bibinfo {pages}
  {1566} (\bibinfo {year} {1995})}\BibitemShut {NoStop}%
\bibitem [{\citenamefont {Lenoir}\ \emph {et~al.}(1996)\citenamefont {Lenoir},
  \citenamefont {Cassart}, \citenamefont {Michenaud}, \citenamefont
  {Scherrer},\ and\ \citenamefont {Scherrer}}]{Lenoir_1996_JPCS}%
  \BibitemOpen
  \bibfield  {author} {\bibinfo {author} {\bibfnamefont {B.}~\bibnamefont
  {Lenoir}}, \bibinfo {author} {\bibfnamefont {M.}~\bibnamefont {Cassart}},
  \bibinfo {author} {\bibfnamefont {J.~P.}\ \bibnamefont {Michenaud}}, \bibinfo
  {author} {\bibfnamefont {H.}~\bibnamefont {Scherrer}}, \ and\ \bibinfo
  {author} {\bibfnamefont {S.}~\bibnamefont {Scherrer}},\ }\href@noop {}
  {\bibfield  {journal} {\bibinfo  {journal} {Journal of Physics and Chemistry
  of Solids}\ }\textbf {\bibinfo {volume} {57}},\ \bibinfo {pages} {89}
  (\bibinfo {year} {1996})}\BibitemShut {NoStop}%
\bibitem [{\citenamefont {Koroteev}\ \emph {et~al.}(2004)\citenamefont
  {Koroteev}, \citenamefont {Bihlmayer}, \citenamefont {Gayone}, \citenamefont
  {Chulkov}, \citenamefont {Blugel}, \citenamefont {Echenique},\ and\
  \citenamefont {Hofmann}}]{Koroteev_2004_PRL}%
  \BibitemOpen
  \bibfield  {author} {\bibinfo {author} {\bibfnamefont {Y.~M.}\ \bibnamefont
  {Koroteev}}, \bibinfo {author} {\bibfnamefont {G.}~\bibnamefont {Bihlmayer}},
  \bibinfo {author} {\bibfnamefont {J.~E.}\ \bibnamefont {Gayone}}, \bibinfo
  {author} {\bibfnamefont {E.~V.}\ \bibnamefont {Chulkov}}, \bibinfo {author}
  {\bibfnamefont {S.}~\bibnamefont {Blugel}}, \bibinfo {author} {\bibfnamefont
  {P.~M.}\ \bibnamefont {Echenique}}, \ and\ \bibinfo {author} {\bibfnamefont
  {P.}~\bibnamefont {Hofmann}},\ }\href@noop {} {\bibfield  {journal} {\bibinfo
   {journal} {Physical Review Letters}\ }\textbf {\bibinfo {volume} {93}},\
  \bibinfo {pages} {046403} (\bibinfo {year} {2004})}\BibitemShut {NoStop}%
\bibitem [{\citenamefont {Marcano}\ \emph {et~al.}(2010)\citenamefont
  {Marcano}, \citenamefont {Sangiao}, \citenamefont {Magen}, \citenamefont
  {Morellon}, \citenamefont {Ibarra}, \citenamefont {Plaza}, \citenamefont
  {Perez},\ and\ \citenamefont {Teresa}}]{Marcano_2010_PRB}%
  \BibitemOpen
  \bibfield  {author} {\bibinfo {author} {\bibfnamefont {N.}~\bibnamefont
  {Marcano}}, \bibinfo {author} {\bibfnamefont {S.}~\bibnamefont {Sangiao}},
  \bibinfo {author} {\bibfnamefont {C.}~\bibnamefont {Magen}}, \bibinfo
  {author} {\bibfnamefont {L.}~\bibnamefont {Morellon}}, \bibinfo {author}
  {\bibfnamefont {M.~R.}\ \bibnamefont {Ibarra}}, \bibinfo {author}
  {\bibfnamefont {M.}~\bibnamefont {Plaza}}, \bibinfo {author} {\bibfnamefont
  {L.}~\bibnamefont {Perez}}, \ and\ \bibinfo {author} {\bibfnamefont
  {J.~M.~D.}\ \bibnamefont {Teresa}},\ }\href@noop {} {\bibfield  {journal}
  {\bibinfo  {journal} {Physical Review B}\ }\textbf {\bibinfo {volume} {82}},\
  \bibinfo {pages} {125326} (\bibinfo {year} {2010})}\BibitemShut {NoStop}%
\bibitem [{\citenamefont {Missana}\ and\ \citenamefont
  {Afonso}(1996)}]{Missana_1996_APA}%
  \BibitemOpen
  \bibfield  {author} {\bibinfo {author} {\bibfnamefont {T.}~\bibnamefont
  {Missana}}\ and\ \bibinfo {author} {\bibfnamefont {C.~N.}\ \bibnamefont
  {Afonso}},\ }\href@noop {} {\bibfield  {journal} {\bibinfo  {journal}
  {Applied Physics A}\ }\textbf {\bibinfo {volume} {62}},\ \bibinfo {pages}
  {513} (\bibinfo {year} {1996})}\BibitemShut {NoStop}%
\bibitem [{\citenamefont {Ast}\ \emph {et~al.}(2007)\citenamefont {Ast},
  \citenamefont {Henk}, \citenamefont {Ernst}, \citenamefont {Moreschini},
  \citenamefont {Falub}, \citenamefont {Pacil\'e}, \citenamefont {Bruno},
  \citenamefont {Kern},\ and\ \citenamefont {Grioni}}]{Ast_2007_PRL}%
  \BibitemOpen
  \bibfield  {author} {\bibinfo {author} {\bibfnamefont {C.~R.}\ \bibnamefont
  {Ast}}, \bibinfo {author} {\bibfnamefont {J.}~\bibnamefont {Henk}}, \bibinfo
  {author} {\bibfnamefont {A.}~\bibnamefont {Ernst}}, \bibinfo {author}
  {\bibfnamefont {L.}~\bibnamefont {Moreschini}}, \bibinfo {author}
  {\bibfnamefont {M.~C.}\ \bibnamefont {Falub}}, \bibinfo {author}
  {\bibfnamefont {D.}~\bibnamefont {Pacil\'e}}, \bibinfo {author}
  {\bibfnamefont {P.}~\bibnamefont {Bruno}}, \bibinfo {author} {\bibfnamefont
  {K.}~\bibnamefont {Kern}}, \ and\ \bibinfo {author} {\bibfnamefont
  {M.}~\bibnamefont {Grioni}},\ }\href {\doibase 10.1103/PhysRevLett.98.186807}
  {\bibfield  {journal} {\bibinfo  {journal} {Phys. Rev. Lett.}\ }\textbf
  {\bibinfo {volume} {98}},\ \bibinfo {pages} {186807} (\bibinfo {year}
  {2007})}\BibitemShut {NoStop}%
\bibitem [{\citenamefont {Niimi}\ \emph {et~al.}(2012)\citenamefont {Niimi},
  \citenamefont {Kawanishi}, \citenamefont {Wei}, \citenamefont {Deranlot},
  \citenamefont {Yang}, \citenamefont {Chshiev}, \citenamefont {Valet},
  \citenamefont {Fert},\ and\ \citenamefont {Otani}}]{Niimi_2012_PRL}%
  \BibitemOpen
  \bibfield  {author} {\bibinfo {author} {\bibfnamefont {Y.}~\bibnamefont
  {Niimi}}, \bibinfo {author} {\bibfnamefont {Y.}~\bibnamefont {Kawanishi}},
  \bibinfo {author} {\bibfnamefont {D.~H.}\ \bibnamefont {Wei}}, \bibinfo
  {author} {\bibfnamefont {C.}~\bibnamefont {Deranlot}}, \bibinfo {author}
  {\bibfnamefont {H.~X.}\ \bibnamefont {Yang}}, \bibinfo {author}
  {\bibfnamefont {M.}~\bibnamefont {Chshiev}}, \bibinfo {author} {\bibfnamefont
  {T.}~\bibnamefont {Valet}}, \bibinfo {author} {\bibfnamefont
  {A.}~\bibnamefont {Fert}}, \ and\ \bibinfo {author} {\bibfnamefont
  {Y.}~\bibnamefont {Otani}},\ }\href {\doibase 10.1103/PhysRevLett.109.156602}
  {\bibfield  {journal} {\bibinfo  {journal} {Phys. Rev. Lett.}\ }\textbf
  {\bibinfo {volume} {109}},\ \bibinfo {pages} {156602} (\bibinfo {year}
  {2012})}\BibitemShut {NoStop}%
\end{thebibliography}%

\end{document}